\documentclass[aps,prl,twocolumn,superscriptaddress]{revtex4-1}\usepackage[utf8]{inputenc}
\usepackage{amsmath}
\usepackage{amssymb}
\usepackage{epsfig}
\usepackage{graphicx,url}
\usepackage{bm}
\usepackage{mathrsfs}
\usepackage{color}
\usepackage[top=2cm, bottom=2.5cm, left=2cm, right=2cm]{geometry}
\usepackage{mathtools}
\usepackage{empheq}
\usepackage{physics}

\newcommand{\revL}{\text{\reflectbox{$\mathcal{L}$}}}
\newcommand{\bos}{\boldsymbol}

\begin{document}

\title{On the dynamics of reaction coordinates in classical, time-dependent, many-body processes}

\author{Hugues Meyer}
\affiliation{\it Physikalisches Institut, Albert-Ludwigs-Universit\"{a}t,  79104 Freiburg, Germany}
\affiliation{Research Unit in Engineering Science, Universit\'{e} du Luxembourg,\\  L-4364 Esch-sur-Alzette, Luxembourg}
\author{Thomas Voigtmann}
\affiliation{Institut f\"{u}r Materialphysik im Weltraum,
	Deutsches Zentrum f\"{u}r Luft- und Raumfahrt (DLR),\\ 51170 K\"{o}ln, Germany}
\affiliation{Department of Physics, Heinrich Heine University, Universit\"{a}tsstra\ss e 1, 40225 D\"{u}sseldorf, Germany}
\author{Tanja Schilling}
\affiliation{\it Physikalisches Institut, Albert-Ludwigs-Universit\"{a}t,  79104 Freiburg, Germany}

\date{\today}

\begin{abstract}
Complex microscopic many-body processes are often interpreted in terms of so-called ``reaction coordinates'', i.e.~in terms of the evolution of a small set of coarse-grained observables. A rigorous method to produce the equation of motion of such observables is to use projection operator techniques, which split the dynamics of the observables into a main contribution and a marginal one. The basis of any derivation in this framework is the classical (or quantum) Heisenberg equation for an observable. If the Hamiltonian of the underlying microscopic dynamics and the observable under study do not explicitly depend on time, this equation is obtained by a straight-forward derivation. However, the problem is more complicated if one considers Hamiltonians which depend on time explicitly as e.g.~in systems under external driving, or if the observable of interest has an explicit dependence on time. We use an analogy to fluid dynamics to derive the classical Heisenberg picture and then apply a projection operator formalism to derive the non-stationary generalized Langevin equation for a coarse-grained variable. We show, in particular, that the results presented for time-independent Hamiltonians and observables in J.~Chem.~Phys.~{\bf 147}, 214110 (2017) can be generalized to the time-dependent case.
\end{abstract}

\maketitle

\section{Introduction}

When studying a complex many-body system, it is often a pragmatic choice to describe a process of interest in terms of the evolution of a small set of relevant observables (``reaction coordinates''), which capture the main features of the process. In molecular biophysics this view is commonly adopted: for example, protein folding, amyloid fiber formation or DNA mechanics are described in terms of the size, shape and structure of the complex molecules involved, while the degrees of freedom of the surrounding water as well as detailed information on the molecular structure are ``averaged over''\cite{Peters2016, Rohrdanz13, Sittel18}. Similarly, chemical reactions and phase transitions are often described in terms of reaction coordinates. In the context of computer simulation this approach is particularly useful. To run a full molecular dynamics simulation of a complex many-body system is often computationally very demanding, therefore it is convenient if one can systematically integrate out degrees of freedom to obtain a self-consistent equation of motion for the observables of interest.

The derivation of such equations from first principles has been discussed over many decades, starting with Langevin in 1908 \cite{Lemons1998} and the introduction of his famous equation, then followed by various resourceful theoretical works, in particular the development of projection operator techniques and the Mori-Zwanzig formalisms \cite{Zwanzig1961a, Mori1965a}. These techniques have first been introduced in equilibrium physics, and progressively extended to non-equilibrium systems \cite{Kawasaki1973, Furukawa1979, Latz2002, Fuchs2009}. In this context, we have recently derived a non-stationary version of the Generalized Langevin Equation as a general structure for the evolution of an arbitrary phase-space function in a non-stationary process \cite{Meyer2017b}. One central hypothesis of the derivation in that paper was to work with a time-independent Liouvillian, i.e.~time-independent microscopic equations of motion. 

Here we show, by two different methods, how to extend the results presented in ref.~\cite{Meyer2017b} to explicitly time-dependent microscopic dynamics as well as explicitly time-dependent observables. First, we make use of a result derived by Holians and Evans in ref.~\cite{Holian1985} on the evolution of phase-space observables, and we derive a general fluctuation-dissipation-like relation that relates the memory kernel with the fluctuating contribution to the Generalized Langevin Equation. Second, we introduce a method that is inspired by fluid dynamics. In the case of an observable that explicitly depends on time, we first show that adding a temporal degree of freedom to phase-space can effectively remove this inconvenient dependence. We then apply this method to time-dependent microscopic dynamics in such a way that the resulting equation of motion can be written using a propagator that has an exponential form. We show how to compute averages in this augmented phase-space and we show the equivalence with the derivation in the standard phase-space.

\section{Projection operator techniques in time-independent microscopic dynamics}

We study a dynamical process, defined as the evolution in time of a large set of degrees of freedom $\bos{\gamma}$ (e.g.~the positions $\bos{q}$ and momenta $\bos{p}$ of a system of many particles). We work on the ensemble level, i.e.~we consider infinitely many copies of the system initialized at time $s$ with some distribution of states $\rho(\bos{\gamma},t=s)$, which does not need to be stationary, i.e.~we could initialize the ensemble out of equilibrium. Assume now that one is interested in the time-evolution of an observable $A$ on this ensemble. In classical statistical mechanics, as in quantum mechanics, there are two approaches: Either one works in the Schrödinger picture, which consists in studying a phase-space distribution $\rho(\bos{\gamma}, t)$ that evolves in time and a fixed observable field $\mathbb{A}(\bos{\gamma})$. Or one works in the Heisenberg picture, which refers to the point of view in which one follows the evolution of the observable $A(\bos{\gamma},s ;t) = \mathbb{A}(\bos{\Gamma}(\bos{\gamma},s;t))$ along each particular trajectory, where $\bos{\Gamma}(\bos{\gamma},s;t)$ denotes the point in phase-space reached at time $t$ given that the trajectory was located at the position $\bos{\gamma}$ at time $s$, and then one averages over the initial distribution. The names of these two complementary pictures are obviously borrowed from quantum mechanics but the idea is similar: in the first picture the density is time-evolved whereas in the other the observable is. For systems in which the microscopic propagator does not explicitly depend on time, one usually writes
\begin{align}
\label{iL_rho}
\frac{\partial}{\partial t}  \rho(\bos{\gamma},t) + i\revL\rho(\bos{\gamma},t) &= 0 \\
\label{iL_A}
\frac{d}{d t}A(\bos{\gamma},s;t) &= i\mathcal{L}A(\bos{\gamma},s;t)
\end{align}
where 
\begin{align}
	i\revL \equiv& \frac{\partial}{\partial \bos{\gamma}} \cdot \left[ \dot{\bos{\gamma}} \cdots \right]\\
	i\mathcal{L} \equiv& \dot{\bos{\gamma}}\cdot \frac{\partial}{\partial \bos{\gamma}}
\end{align}
is the Liouvillian operator and $\dot{\bos{\gamma}}$ is a vector flow field determined by the microscopic equations of motion. (The case in which $\dot{\bos{\gamma}}$ does not explicitly depend on time is what we refer to as time-independent microscopic dynamics). The vector field $\dot{\bos{\gamma}}$ then defines a flow $\bos{\Gamma}(\bos{\gamma},s;t)$ as the solution of the initial-value problem $(d/dt){\bos{x}}_s(t)=\dot{\bos\gamma}(\bos{x}_s(t))$ with $\bos{x}_s(s)=\bos{\gamma}$, where the flow is defined as the mapping $\bos{\Gamma}(\bos{\gamma},s;t)=\bos{x}(t)$. Note that $(d/dt)\bos{x}_s(t+\tau)=\dot{\bos\gamma}_s(\bos{x}(t+\tau))$, but also $(d/dt)\bos{x}_\tau(t)\circ\bos{x}_s(\tau)=\dot{\bos\gamma}(\bos{x}_\tau(t)\circ\bos{x}_s(\tau))$, where we set $\bos{x}_\tau(t)\circ\bos{x}_s(\tau)\equiv\bos{\Gamma}(\bos{\Gamma}(\bos{\gamma},s;\tau),\tau,t)$. Hence, $\bos{x}_s(t+\tau)$ and $\bos{x}_\tau(t)\circ\bos{x}_s(\tau)$ both obey the same first-order differential equation with the same initial value at $t=s$. They are thus identical. This identity is at the basis of writing $\bos{\Gamma}$ and related quantities in terms of exponential operators. 

Equation (\ref{iL_rho}) is known as Liouville's theorem, while equation (\ref{iL_A}) is easily derived from a chain rule. Their formal solutions are given by  
\begin{align}
  \rho(\bos{\gamma},t) &= e^{-i\revL (t-s)}\rho(\bos{\gamma},s) \\
  \label{eqn:propagate}
A(\bos{\gamma},s;t) &= e^{i\mathcal{L}(t-s)}\mathbb{A}(\bos{\gamma})
\end{align}
where we have used the fact that the phase-space fields $A(\bos{\gamma},s;s)$ and $\mathbb{A}(\bos{\gamma})$ are equal. By defining an inner product on phase-space as
\begin{equation}
\label{scalar}
\left[ X,Y \right] = \int d\bos{\gamma} X^{*}(\bos{\gamma}) Y(\bos{\gamma})
\end{equation}
where the star stands for the complex conjugate, we can write the average of $A(t)$ in the Heisenberg picture and the Schrödinger picture.(Note that we assume in the following that we always deal with functions for which this integral exists.) In the Schrödinger picture we obtain
\begin{equation}
\left\langle A(t) \right\rangle = \left[ \rho(\bos{\gamma},t), \mathbb{A}(\bos{\gamma}) \right] = \left[ e^{-i\revL (t-s)}\rho(\bos{\gamma},s), \mathbb{A}(\bos{\gamma}) \right]
\end{equation}
while in the Heisenberg picture we obtain
\begin{equation}
\left\langle A(t) \right\rangle = \left[ \rho(\bos{\gamma},s), A(\bos{\gamma},s;t) \right] = \left[ \rho(\bos{\gamma},s), e^{i\mathcal{L}(t-s)}\mathbb{A}(\bos{\gamma}) \right]
\end{equation}
The definition (\ref{scalar}) of the inner product is such that $i\mathcal{L}$ and $i\revL$ are anti-adjoints, i.e. $(i\mathcal{L})^{\dagger} = -i\revL$. This ensures the equivalence of the averages. Note that in conservative systems (i.e.~such that $\frac{\partial }{\partial \bos{\gamma}} \cdot \dot{\bos{\gamma}} = 0$), we have $i\revL = i\mathcal{L}$, which becomes thus anti-self-adjoint. This is the case of autonomous Hamiltonian dynamics for instance.

Projection operator techniques are commonly used to study the time-evolution of a dynamical variable $A(\bos{\gamma},t)$ (we will omit $\bos{\gamma}$ in the rest of this paragraph). The main idea is to split the dynamics of the observable in the Heisenberg picture into a main {\it parallel} contribution and a remaining {\it orthogonal} part. The standard derivation \cite{Grabert1982} proceeds as follows:

Let us define an operator $P_{t}$, that can be time-dependent, which acts in the space of phase-space functions and is such that for any pair of times $t$ and $t'$ the following identity is satisfied:
\begin{equation}
\label{ptptp}
	P_{t'}P_{t} = P_{t}
\end{equation}
This property defines $P_{t}$ as a projection operator: it retrieves a certain fixed contribution from the variable it is applied to. By decomposing the identity operator as $\mathbb{I} = P_{t} + (\mathbb{I} - P_{t})$ and inserting it in the following way in eqn.~\ref{eqn:propagate}
\begin{equation}
\label{split_dynamics}
	\frac{dA_{t}}{dt} = e^{i\mathcal{L}t} \left[ P_{t} + Q_{t} \right] i\mathcal{L} A_{0}
\end{equation}
where $A_{t} = A(\bos{\gamma},0 ;t)$ is the value of the observable $A$ at time $t$, and $Q_{t} = 1 - P_{t}$, it is possible to rewrite its time-evolution as \cite{Meyer2017b}
\begin{align}
\label{EOM_A}
\frac{dA_{t}}{dt} =& e^{i\mathcal{L}t}P_{t}i\mathcal{L}A_{0} \nonumber \\
&+ \int_{t'}^{t}{d\tau e^{i\mathcal{L}\tau}P_{\tau} \left[i\mathcal{L} - \dot{P}_{\tau} \right] Q_{\tau} G_{\tau,t}i\mathcal{L}A_{0}} \nonumber \\
&+ e^{i\mathcal{L}t'}Q_{t'}G_{t',t}i\mathcal{L}A_{0}
\end{align}
where $t'$ is an arbitrary reference time and $G_{\tau,t}$ is a negatively time-ordered exponential defined as
\begin{align}
\label{time_order}
G_{t',t} =& \exp_{-}\left\{\int_{t'}^{t} dt'' i\mathcal LQ_{t''} \right\} \nonumber \\
= 1 + &\sum_{n=1}^{\infty} \int_{t'}^{t} dt_{1} \int_{t'}^{t_{1}} dt_{2} \cdots \int_{t'}^{t_{n-1}} dt_{n} i\mathcal LQ_{t_{n}} \cdots i\mathcal LQ_{t_{1}}
\end{align}
The next step is then to specify the projection operator in order to obtain a structural form for the equation of motion of $A(t)$. In general terms, we can write $P_{t}$ as an operator that projects onto a set of relevant phase-space functions $\left\{ F_{i} \right\}$ , i.e. its action on any phase-space function $X$ is
\begin{equation}
	P_{t}X = \sum_{i} \mathcal{O}_{t}(X,F_{i})F_{i}
\end{equation}
where $\mathcal{O}_{t}(X,F_{i})$ is a scalar prefactor, which to choose one has a certain freedom. In most cases the variable under study $A$ is itself the relevant variable that one projects onto. In ref.~\cite{Meyer2017b} we have shown that, if one intends to study the auto-correlation function $C(t,t') = \left\langle A_{t}A^{*}_{t'} \right\rangle$, one would choose 
\begin{equation}
	P_{t}X = \left\langle A_{t}^{*} X_{t} \right\rangle \left\langle \left| A_{t} \right|^{2} \right\rangle^{-1}A
\end{equation}
in order to obtain the following equations of motion for $A_{t}$ and $C(t,t')$:
\begin{align}
\label{EOM_A}
\frac{dA_{t}}{dt} =&  \omega(t)A_{t} + \int_{t'}^{t}{d\tau K(t,\tau)A_{\tau} } + \eta_{t'}(t',t) \\
\label{EOM_C}
\frac{dC(t',t)}{dt} =& \omega_{1}(t)C(t',t) + \int_{t'}^{t}{d\tau K(t,\tau) C(t',\tau)} 
\end{align}
where we have defined the following quantities
\begin{align}
\label{def_omega}
\omega_{1}(t) &= \left \langle A_{t}^{*} i\mathcal{L}A_{t}   \right\rangle \left\langle \left| A_{t} \right|^{2} \right\rangle^{-1} \\
\label{def_K}
K(t,\tau) &= \left\langle A_{\tau}^{*} e^{i\mathcal{L}\tau}  \left[ i\mathcal{L} - \dot{P}_{\tau} \right]Q_{\tau} G_{\tau,t}i\mathcal{L}A_{0}  \right\rangle \left\langle \left| A_{t} \right|^{2} \right\rangle^{-1} \\
\label{def_eta}
\eta_{t'}(t',t) &= e^{i\mathcal{L}t'}Q_{t'}G_{t',t}i\mathcal{L}A_{0}
\end{align}

However, this derivation is invalid if eqn.~(\ref{iL_A}) does not hold. In particular, care is required in the case of a time-dependent Liouvillian operator, i.e.~a system in which the flow field in phase-space is not constant in time. To overcome the problem, we have to adapt equations (\ref{iL_rho}) and (\ref{iL_A}) to the case of explicitly time-dependent microscopic dynamics. In the classical case as well as in the quantum mechanical one, the Schr\"odinger picture is easily generalized to the time-dependent dynamics. In fact, the conservation of probability, when written with all the space and time dependences, still reads
\begin{equation}
\label{liouville_theorem_timedep}
\frac{\partial \rho(\bos{\gamma},t)}{\partial t} +  \frac{\partial}{\partial \bos{\gamma}} \cdot \left[\rho(\bos{\gamma},t)\dot{\bos{\gamma}}(\bos{\gamma},t)\right]= 0
\end{equation}
In words, as the total probability of finding the system in phase-space is obviously conserved and as the trajectories in phase-space are continuous, the instantaneous change in the probability density at a local point $\bos{\gamma}$ is due to the incoming and outgoing probability fluxes that pass by the point $\bos{\gamma}$. At time $t$, the total flux is equal to $\rho(\bos{\gamma},t)\dot{\bos{\gamma}}(\bos{\gamma},t)$ hence the structure of the balance equation (\ref{liouville_theorem_timedep}). Defining the time-dependent generator
\begin{equation}
\label{Liouville_adjoint_timedep}
i\revL_{t} \equiv \frac{\partial}{\partial \bos{\gamma}} \cdot\left[ \dot{\bos{\gamma}}(\bos{\gamma},t) \cdots \right]
\end{equation}
we conclude that eqn.~(\ref{iL_rho}) remains valid for classical, time-dependent microscopic dynamics when replacing $i\revL$ by $i\revL_{t}$. It can formally be integrated for $\rho$ because its structure remains the one of a partial differential equation in $\rho(\bos{\gamma}, t)$. We obtain
\begin{equation}
\label{rho_texp}
\rho(\bos{\gamma},t) = \exp_{+}\left\{-\int_{s}^{t}i\revL_{t'}dt' \right\} \rho(\bos{\gamma},s)
\end{equation} 
where $\exp_{+}$ denotes the positively time-ordered exponential. The time-dependent average $\left\langle A(t) \right\rangle$ would then be written as
\begin{equation}
\label{ave_A_schrodinger}
\left\langle A(t) \right\rangle =  \left[ \exp_{+}\left\{-\int_{s}^{t}i\revL_{t'}dt'\right\} \rho(\bos{\gamma},s), \mathbb{A}(\bos{\gamma}) \right]
\end{equation}

On the other hand, the time-evolution of an observable in the Heisenberg picture in the case of time-dependent Liouvillian operators requires more care. A time-dependent vector field $\dot{\bos\gamma}$ does not readily define a flow in the same way a time-independent vector field does.
Using the notation defined above, we would get
$(d/dt)\bos{x}_s(t+\tau)=\dot{\bos{\gamma}}_{t+\tau}(\bos{x}_s(t+\tau))$ but
$(d/dt)\bos{x}_\tau(t)\circ\bos{x}_s(\tau)=\dot{\bos{\gamma}}_t(\bos{x}_\tau(t)\circ\bos{x}_s(\tau))$, which is not identical. Hence it does not make sense to write $\bos{x}$ or related quantities as simple exponentials. 

As Holian and Evans have shown \cite{Holian1985}, it is not correct to simply replace $i\mathcal{L}$ by $i\mathcal{L}_{t}$ in (\ref{iL_A}) in the case of time-dependent microscopic dynamics. That would lead to serious inconsistencies with eqn.~(\ref{liouville_theorem_timedep}). The problem can be tackled in the following way: Let us denote as $U_{s,t}$ the operator that advances the implicitly time-dependent phase variables from a previous time $s$ to the current time $t$, along the trajectory, i.e.
\begin{equation}
\bos{\Gamma}\left( \bos{\gamma}, s; t \right) = U_{t',t}\bos{\Gamma}\left( \bos{\gamma},s; t' \right)
\end{equation}
In particular, $U_{t',t}$ is defined such that $\dot{\bos{\gamma}}\left(\bos{\Gamma}\left( \bos{\gamma},s; t \right), t\right) = U_{s,t}\dot{\bos{\gamma}}( \bos{\gamma},t )$, and for any pair of observables $X(\bos{\gamma},s; t)$ and $Y(\bos{\gamma},s; t)$ we have $X(\bos{\gamma},s;t)Y(\bos{\gamma},s;t) = U_{s,t}\left[ X(\bos{\gamma},s;s)Y(\bos{\gamma},s;s) \right] = U_{s,t}\left[ X(\bos{\gamma},s;s)\right] U_{s,t}\left[Y(\bos{\gamma},s;s) \right]$. Finally, we emphasize the relation $\mathbb{A}\left(\bos{\Gamma}(\bos{\gamma},s;t)\right) = U_{s,t}\mathbb{A}(\bos{\gamma})$. Differenciating this identity with respect to $t$ and using the correspondance between $d_{t}\bos{\Gamma}$ and $\dot{\bos{\gamma}}$ (that we will explain in more details later) we obtain 
\begin{equation}
\label{Ust_dt}
\frac{d }{d t} U_{s,t}\mathbb{A}(\bos{\gamma}) =  U_{s,t}\left[\dot{\bos{\gamma}}\left( \bos{\gamma},t \right) \cdot \frac{\partial}{\partial \bos{\gamma}}  \mathbb{A}\left(\bos{\gamma} \right)\right]
\end{equation}
As this equation holds for any obervable field $\mathbb{A}$, by defining $i\mathcal{L}_{t} = \dot{\bos{\gamma}}\left( \bos{\gamma},t \right) \cdot \frac{\partial}{\partial \bos{\gamma}}$, one obtains 
\begin{equation}
\frac{d }{dt} U_{s,t} = U_{s,t}i\mathcal{L}_{t}
\end{equation}
This equation is solved using a negatively time-ordered exponential, which then allows us to write
\begin{equation}
A(\bos{\gamma},s;t) = \exp_{-}\left\{\int_{s}^{t}i\mathcal{L}_{t'}dt' \right\} A(\bos{\gamma},s;s)
\end{equation}
This result is perfectly consistent with the one previously obtained for $\rho(\bos{\gamma},t)$, eqn.~(\ref{rho_texp}), because of the adjointness identity
\begin{equation}
\left[ \exp_{-}\left\{\int_{s}^{t}i\mathcal{L}_{t'}dt' \right\}\right]^{\dagger} = \exp_{+}\left\{-\int_{s}^{t}i\revL_{t'}dt' \right\}
\end{equation}
Writing the time-dependent average $\left\langle A(t) \right\rangle$ in the Heisenberg picture using this result would coincide with the one shown in eqn.~(\ref{ave_A_schrodinger}). This result is very convenient in order to apply the projection operator formalism. In fact we have
\begin{equation}
\frac{d A_{t}}{dt} = \exp_{-}\left\{\int_{0}^{t}i\mathcal{L}_{t'}dt' \right\} i\mathcal{L}_{t} A_{0}
\end{equation}
where $A_{t}= A(\bos{\gamma}, 0; t)$ again. We can thus again insert a time-dependent projection operator $P_{t}$ as 
\begin{equation}
\frac{d A_{t}}{dt} = \exp_{-}\left\{\int_{0}^{t}i\mathcal{L}_{t'}dt' \right\} \left[ P_{t} i\mathcal{L}_{t} A_{0} +  Q_{t} i\mathcal{L}_{t} A_{0}\right]
\end{equation}
Following Grabert \cite{Grabert1982}, one writes a differential equation for the quantity $Z_{t}=\exp_-[\int_{0}^{t} i\mathcal L_{t'} dt'] Q_{t}$:
\begin{align}
 \dot{Z}_{t} =& Z_{t}i\mathcal L_{t}Q_{t}+ \exp_-\left\{\int_{0}^{t} i\mathcal L_{t'}dt'\right\}
  \mathcal P_t\left[i\mathcal L_{t}-\dot{P}_t \right] Q_t
\end{align}
from which follows
\begin{align}
  Z_{t} =& Z_{t'} G_{t',t} \nonumber \\
  +& \int_{t'}^t d\tau  \exp_-\left\{ \int_{0}^{\tau} i\mathcal L_{t''} dt'' \right\}  P_{\tau}\left[i\mathcal L_{\tau}-\dot{P}_{\tau}\right] Q_\tau G_{\tau,t}
\end{align}
where $G_{t',t}$ is defined in eqn.~(\ref{time_order}) and $t'$ is an arbitrary reference time. With this, we can apply the same machinery as in ref.~\cite{Meyer2017b} and show that equations (\ref{EOM_A}) and (\ref{EOM_C}) are still valid, provided that $e^{i\mathcal{L}t}$ is replaced by $\exp_-\left\{ \int_{0}^{t} i\mathcal L_{t'} dt' \right\}$ in $\omega_{1}$, $K$ and $\eta$.

This general result is interesting, because it shows that the structure of a Generalized Langevin Equation is quite robust. The term ``time-dependent microscopic dynamics'' encompasses time-dependent Hamiltonian processes as well as other non-conservative processes. We have proven that a constant structure remains valid in all these cases and thus the tools and methods to analyse data in terms of Generalized Langevin Equations \cite{Yoshimoto2017, Jung2017, Jung2018, Zaccone2018, Lange2006, Makroglou1992} can be systematically applied.

\subsection*{A fluctuation-dissipation-like relation}
The definition of the projection operator $P$ allows us to derive a relation which links the friction kernel and the autocorrelation of the fluctuating force
\begin{equation}
\label{fluc_diss}
K(t,t') \left\langle |A_{t'}|^{2} \right\rangle = -\left\langle \eta_{t'}^{*}(t',t')\eta_{t'}(t',t) \right\rangle
\end{equation}
This identity holds, because the proof shown in \cite{Meyer2017b} remains valid in the case of time-dependent microscopic dynamics. In equilibrium physics, this structure is known as a fluctuation-dissipation theorem. Out of equilibrium, we cannot interpret the effect of the kernel as dissipation in general. However, here we show that the kernel is fundamentally related to the autocorrelation of the fluctuating force. As the kernel( together with the drift $\omega(t)$) is the main function that controls the auto-correlation of the observable $A$ (see \ref{EOM_C}) this FDT-like relation therefore links the autocorrelation of the observable with the fluctuating force. However, eqn.~(\ref{fluc_diss}) involves the auto-correlation function of the fluctuating force $\eta$ between an arbitrary time $t$ and the initial (or 'reference') time $t'$. It would be more useful to derive a similar type of relation for the auto-correlation function of $\eta$ between two arbitrary times $t_{1}$ and $t_{2}$, i.e. for $\left\langle \eta_{t'}^{*}(t',t_{1})\eta_{t'}(t',t_{2}) \right\rangle$. This would for instance help if one aims to numerically generate random realisations of the fluctuating force $\eta$ according to its correct distribution. To do this, we go back to eqn.~(\ref{EOM_A}) and we take its derivative with respect to $t'$ reading
\begin{equation}
0 = - K(t,t') A_{t'} + \frac{\partial \eta_{t'}(t',t)}{\partial t'}
\end{equation}
which we can integrate to find
\begin{equation}
\label{solution_eta}
\eta_{t'}(t',t) = \eta_{t_{1}}(t_{1},t) - \int_{t'}^{t_{1}} d\tau K(t,\tau) A_{\tau} 
\end{equation}
where $t_{1}$ is an arbitrary reference time. Now, we calculate $\left\langle \eta_{t'}(t',t_{1})^{*} \eta_{t'}(t',t_{2}) \right\rangle$ by multiplying (\ref{solution_eta}) by itself, once with $t=t_{1}$ and once with $t=t_{2}$ :
\begin{align}
\label{long_etaeta}
\left\langle \eta_{t'}(t',t_{1})^{*}\eta_{t'}(t',t_{2}) \right\rangle = -&K(t_{2},t_{1})\left\langle |A_{t_{1}}|^{2} \right\rangle \nonumber \\
+ \int_{t'}^{t_{1}} d\tau \int_{t'}^{t_{1}} d\tau' &K^{*}(t_{1},\tau) K(t_{2},\tau') \left\langle A_{\tau}^{*} A_{\tau'}  \right\rangle \nonumber \\
- \int_{t'}^{t_{1}} d\tau &K^{*}(t_{1},\tau) \left\langle A_{\tau}^{*} \eta_{t_{1}}(t_{1},t_{2})  \right\rangle \nonumber \\
- \int_{t'}^{t_{1}} d\tau &K(t_{2},\tau) \left\langle A_{\tau} \eta_{t_{1}}^{*}(t_{1},t_{1})  \right\rangle
\end{align}
where we have already used (\ref{fluc_diss}) to rewrite the first term in the right-hand side. Let us now simplify the last two terms involving $ \left\langle A_{\tau}^{*} \eta_{t_{1}}(t_{1},t_{2})  \right\rangle$ and $\left\langle A_{\tau} \eta_{t_{1}}^{*}(t_{1},t_{1}) \right\rangle$. Let us first multiply (\ref{solution_eta}) by $A(\tau)$ and take the average to find
\begin{align*}
\left\langle A_{\tau}^{*}\eta_{t'}(t',t) \right\rangle =& - \int_{t'}^{\tau}d\tau' K(t,\tau') C(\tau',\tau)
\end{align*}
where we have used $\left\langle A_{\tau}^{*}\eta_{\tau}(\tau,t) \right\rangle = 0$ by definition of the projection operator. Therefore we have
\begin{align}
\int_{t'}^{t_{1}} &d\tau K^{*}(t_{1},\tau) \left\langle A_{\tau}^{*} \eta_{t_{1}}(t_{1},t_{2})  \right\rangle \nonumber  \\
&= \int_{t'}^{t_{1}} d\tau \int_{\tau}^{t_{1}} d\tau' K^{*}(t_{1},\tau)  K(t_{2},\tau') C(\tau', \tau)
\end{align}
and 
\begin{align}
\int_{t'}^{t_{1}} &d\tau K(t_{2},\tau) \left\langle A_{\tau} \eta_{t_{1}}^{*}(t_{1},t_{1})  \right\rangle  \nonumber \\
=& \int_{t'}^{t_{1}} d\tau \int_{t'}^{\tau} d\tau' K(t_{2},\tau') K^{*}(t_{1},\tau) C(\tau', \tau)
\end{align}
where we have swapped the order of integration, and relabeled the integration variables . We obtain thus, for real-valued observables
\begin{align}
\int_{t'}^{t_{1}} d\tau &K^{*}(t_{1},\tau) \left\langle A_{\tau}^{*} \eta_{t_{1}}(t_{1},t_{2})  \right\rangle \nonumber \\
+ &\int_{t'}^{t_{1}} d\tau K(t_{2},\tau) \left\langle A_{\tau} \eta_{t_{1}}^{*}(t_{1},t_{1})  \right\rangle \nonumber \\
&= \int_{t'}^{t_{1}} d\tau \int_{t'}^{t_{1}} d\tau' K(t_{2},\tau') K^{*}(t_{1},\tau) C(\tau', \tau)
\end{align}
The last three terms of eqn.~(\ref{long_etaeta}) cancel each other and it finally reads
\begin{equation}
\label{flucdiss_improved}
K(t_{2},t_{1})\left\langle |A_{t_{1}}|^{2} \right\rangle = -\left\langle \eta_{t'}(t',t_{1})\eta_{t'}(t',t_{2}) \right\rangle
\end{equation}
The structure of the ``fluctuation-dissipation-like relation'' still holds true for two arbirtary times, in the case of real-valued observables. Note that the relation is independent from the reference time $t'$ and that it might be a practical tool if one wants to generate trajectories that are solutions of eqn.~(\ref{EOM_A}).

\section*{Treating time as a phase-space coordinate}
Here we introduce a concept, inspired by fluid dynamics, that we consider a useful alternative pathway to the derivation of the Heisenberg equation. This concept also allows us to derive the Generalized Langevin Equation for time-dependent variables, which turns out to be once again identical to the one derived for time-independent observables in ref.~\cite{Meyer2017b}.

\subsection*{The case of time-dependent variables}

Assume now one is interested in a variable that has an explicit dependence on time and one wants to derive the projection operator formalism in order to obtain a Langevin-like equation. The notation for the phase-space field $\mathbb{A}(\bos{\gamma})$ becomes $\mathbb{A}(\bos{\gamma}, t)$ and $A(\bos{\gamma},s;t)$ becomes $A\left( \left[\bos{\gamma},s;t\right] , t \right)$ (where the second time-dependence is an explicit dependence). Both notations are related via $A\left( \left[\bos{\gamma},s;t\right] , t \right) = \mathbb{A}(\bos{\Gamma}(\bos{\gamma},s;t), t)$. In order to apply the method of Holian and Evans in this case, we need to define the operator $U_{s,t}$ such that it also advances the explicit time-dependence of phase-space fields, i.e. $\mathbb{A}(\bos{\Gamma}(\bos{\gamma},s;t), t) = U_{s,t}\mathbb{A}(\bos{\gamma}, s)$. The differentiation with respect to $t$ would then yield
\begin{align}
\frac{d }{d t} U_{s,t}\mathbb{A}(\bos{\gamma}) =   \dot{\bos{\gamma}}\left(\bos{\Gamma}(\bos{\gamma}, s; t), t\right)  \cdot & \frac{\partial \mathbb{A}}{\partial \bos{\gamma}}  \left(\bos{\Gamma}(\bos{\gamma}, s; t), t \right) \nonumber \\
+ & \frac{\partial\mathbb{A}}{\partial t}  \left(\bos{\Gamma}(\bos{\gamma}, s; t), t \right)
\end{align}
In order to recover the convenient structure of eqn.~(\ref{Ust_dt}) that was used for time-independent observables, we can artificially treat the time-dependence of $\mathbb{A}$ as an effective phase-space coordinate and get rid of the partial derivative with respect to time. We hence define an augmented phase-space, which consists in the usual phase-space to which we add an extra temporal dimension. We note the coordinate in this new space 
\begin{equation}
\label{gamma_prime}
\bos{\gamma}' = (\bos{\gamma}, \tau) 
\end{equation}
Along a trajectory, a phase-space particle evolves forward in the $\tau$-direction at a constant rate and formally follows the time-evolution of the process. The flow field in this augmented space is given by
\begin{equation}
\label{gammadot_prime}
\dot{\bos{\gamma}}'(\bos{\gamma}', t) = ( \dot{\bos{\gamma}}(\bos{\gamma}', t), \dot{\tau}(\bos{\gamma}', t)) = ( \dot{\bos{\gamma}}(\bos{\gamma}, \tau), 1)
\end{equation}
In this context $\mathbb{A}(\bos{\gamma}')$ is the observable field in this space, and we can rewrite 
\begin{equation}
\frac{d }{d t} U_{s,t}\mathbb{A}(\bos{\gamma}') =  U_{s,t}\left[\dot{\bos{\gamma}'}\left( \bos{\gamma}',t \right) \cdot \frac{\partial}{\partial \bos{\gamma}}  \mathbb{A}\left(\bos{\gamma}' \right)\right]
\end{equation}
We can then change the definition of the Liouvillian operator in this augmented phase-space accordingly and find 
\begin{equation}
\frac{d }{dt} U_{s,t} = U_{s,t}i\mathcal{L}'_{t}
\end{equation}
The formalism developed in the previous paragraph would then remain valid and the structure of eqn.~(\ref{EOM_A}) would still be valid, together with the FDT-like relation (\ref{flucdiss_improved}). In order to recover the true dependence of the observable $A$, we make sure that the initial augmented phase-space density $\rho'(\bos{\gamma'}, s)$ has a Dirac peak at $\tau = s$, i.e. $\rho(\bos{\gamma'}, s) = \rho(\bos{\gamma}, s)\delta(\tau-s)$. The averages and correlation functions computed in this way are then consistent, such that eqn.~(\ref{EOM_C}) would also hold true. Note that this procedure is in principle valid for any time-dependence in the variable $A$. 

An important example of such time-dependence is the investigation of fluctuations in a non-stationary process. If the time-dependent average $\left\langle A(t) \right\rangle$ of the variable $A$ is not constant, the autocorrelation function $\left\langle A(t) A(t') \right\rangle$ might not be enough to study fluctuations and one may want to study the shifted correlation function 
\begin{equation}
\tilde{C}(t,t') = \left\langle A(t) A(t') \right\rangle - \left\langle A(t) \right\rangle \left\langle A(t')  \right\rangle
\end{equation}
This function can be seen as the true correlation function of the modified variable $\tilde{A}$ whose observable field is defined as
\begin{equation}
\label{modified_A}
\tilde{\mathbb{A}}(\bos{\gamma}, t) = \mathbb{A}(\bos{\gamma}) - \left\langle A(t) \right\rangle
\end{equation}
which is such that $\tilde{C}(t,t') = \left\langle \tilde{A}(t) \tilde{A}(t') \right\rangle$. For this modified variable, the arguments presented above show that the derivation leading to (\ref{EOM_A}) and (\ref{EOM_C}) still holds true and we can hence write
\begin{equation}
\frac{\partial \tilde{C}(t,t')}{\partial t} = \tilde{\omega}(t)\tilde{C}(t,t') + \int_{t'}^{t}d\tau \tilde{K}(t,\tau \tilde{C}(t',\tau)
\end{equation}
where $\tilde{\omega}$ and $\tilde{K}$ are defined smililarly as in (\ref{def_omega}) and (\ref{def_K}) where $A$ is replaced by $\tilde{A}$. Note also that the corresponding fluctuation term $\tilde{\eta}$ has a vanishing average $\left\langle \tilde{\eta}_{t'}(t',t) \right\rangle = 0$ by definition of the variable $\tilde{A}$.  

The concept of augmented phase-space introduced here can now be used to shed a new light on the case of time-dependent microscopic dynamics. 

\subsection*{Reconsidering time-dependent microscopic dynamics}

The approach presented in the previous paragraph can be related to concepts of fluid dynamics in which, as in statistical and quantum mechanics, one has two choices when it comes to describing the motion of an object. One can choose to describe a process in the Euler picture, which consists in looking at the flow of fluid that passes through a fixed point of space. Or one can choose to follow the trajectory of one particle of the fluid throughout the process, the Lagrange picture. An introduction to these concepts is presented in ref.~\cite{Batchelor1999}. The analogy between Hamiltonian mechanics and fluid dynamics has been used successfully multiple times to solve various problems, as shown in ref.~\cite{Salmon1988, Constantin2001, Oseledets1989}. 

To derive the equation of motion, we employ the Lagrange picture, in which one follows the evolution of one particle of ``fluid'' (be it a hydrodynamic volume in real space or a volume in phase-space). The position at time $t$ in that space of a particle that was located at position $\bos{\gamma}$ at time s is denoted by $\bos{\Gamma}(\bos{\gamma}, s; t)$. The change of reference frame with respect to the Euler picture is given by the relation
\begin{equation}
\label{Gamma_gammadot}
\frac{d\bos{\Gamma}(\bos{\gamma},s; t)}{d t} = \dot{\bos{\gamma}}\left(\bos{\Gamma}(\bos{\gamma}, s; t), \tau\right) |_{\tau = t}
\end{equation}
where $\dot{\bos{\gamma}}(\bos{\gamma},t)$ is the flow field at the position $\bos{\gamma}$ at time $t$ and is given by the microscopic equations of motion. We now follow the value of the observable $A$ along a trajectory and denote as $A(\bos{\gamma}, s; t) = \mathbb{A}(\bos{\Gamma}(\bos{\gamma}, s;t))$ its time evolution along a trajectory that was initialized at time $s$ at a position $\bos{\gamma}$ in phase-space. Thus, one has
\begin{align}
A(\bos{\gamma}, s; t+dt) &= \mathbb{A}\left( \bos{\Gamma}\left(\bos{\gamma}, s; t+dt\right) \right) \nonumber \\
&= \mathbb{A}\left(\bos{\Gamma}(\bos{\gamma}, s; t) + \dot{\bos{\gamma}}(\bos{\Gamma}(\bos{\gamma}, s; t),t)dt\right)
\end{align}
i.e.~the chain rule, which yields
\begin{equation}
\label{partial_A_heisenberg}
\frac{d A(\bos{\gamma}, s; t)}{d t} =  \dot{\bos{\gamma}}\left( \bos{\Gamma}(\bos{\gamma}, s; t),t \right) \cdot \frac{\partial \mathbb{A}}{\partial \bos{\gamma}}\left(\bos{\Gamma}(\bos{\gamma}, s; t) \right)
\end{equation}
(Note that the notation $\frac{\partial \mathbb{A}}{\partial \bos{\gamma}}\left(\bos{\Gamma}(\bos{\gamma}, s; t) \right)$ means that we take the derivative of $ \mathbb{A}$ with respect to $\bos{\gamma}$, and we then evaluate it at the point $\bos{\Gamma}(\bos{\gamma}, s; t) $, i.e. $\frac{\partial \mathbb{A}}{\partial \bos{\gamma}} \left(\bos{\Gamma}(\bos{\gamma}, s; t) \right) \neq \frac{d}{d \bos{\gamma}} \left[\mathbb{A}\left(\bos{\Gamma}(\bos{\gamma}, s; t) \right)\right]$. 

We show in the supplemental material that for time-independent dynamics the identity
\begin{equation}
\label{disp_along}
\bos{\Gamma}(\bos{\gamma}+\dot{\bos{\gamma}}(\bos{\gamma})dt, s; t) = \bos{\Gamma}(\bos{\gamma}, s; t+dt)
\end{equation}
is enough to show the validity of the Heisenberg equation. This is due to the fact that the streamlines defined by $\dot{\gamma}$ are constant in time. When written with all the dependences, the corresponding Heisenberg equation reads
\begin{equation}
\label{heisenberg_extensive}
\frac{d A(\bos{\gamma}, s; t)}{d t} =  \dot{\bos{\gamma}}(\bos{\gamma}) \cdot \frac{\partial A(\bos{\gamma}, s; t)}{\partial \bos{\gamma}}  = i\mathcal{L}A(\bos{\gamma},s; t)
\end{equation}
Since eqn.~(\ref{disp_along}) is not valid for time-dependent flow fields $\dot{\gamma}(\bos{\gamma}, t)$, equation (\ref{heisenberg_extensive}) cannot be valid either.  Time-dependent microscopic dynamics is often tackled in very mathematical words, using the tools of topology of symplectic manifolds, which might be hard to read for non-specialists \cite{Wiggins2003, Sardanashvily1998}. Here we will account for the time-dependence of the flow field in the Lagrange picture first, and later in the Euler picture, using hydrodynamics as a didactic tool. As mentioned above, working in a phase-space whose streamlines are fixed in time is intrinsically more convenient. In fact, the very reason that makes the Heisenberg picture valid in time-independent microscopic dynamics is the validity of eqn.~(\ref{disp_along}). 

To achieve this, we use the same technique as for the case of time-dependent observables, namely we add a temporal dimension to phase-space, exactly as defined in eqns.~(\ref{gamma_prime}) and (\ref{gammadot_prime}). The streamlines of this space are constant in time, and the trajectory of a phase-space particle is therefore well-defined in this space. We denote as $\bos{\Gamma}'(\bos{\gamma}', t)$ the position at time $t$ in the augmented phase-space of a trajectory that was initialized at the position $\bos{\gamma}' = (\bos{\gamma}, \tau)$. 
The argument that leads to eqn.~(\ref{heisenberg_extensive}) can now be reproduced in the augmented space. In particular we show
\begin{equation}
\label{augm_jacobian}
\frac{d \bos{\Gamma}'(\bos{\gamma}', s; t)}{d t} =  \frac{\partial \bos{\Gamma}'(\bos{\gamma}', s; t)}{\partial \bos{\gamma}'} \cdot  \dot{\bos{\gamma}'}(\bos{\gamma}') 
\end{equation}
which is the extension of eqn.~(\ref{jacobian_supp}) in the appendix. We now split this into two parts, the phase-space part and the temporal part, and we notice that the equation is valid for any component of $\bos{\Gamma}'$, in particular any phase-space component $\bos{\Gamma}$. Thus, we obtain
\begin{equation}
\label{augm_jacobian}
\frac{d \bos{\Gamma}(\bos{\gamma}', s; t)}{d t} =  \frac{\partial \bos{\Gamma}(\bos{\gamma}', s; t)}{\partial \bos{\gamma}} \cdot  \dot{\bos{\gamma}}(\bos{\gamma}')  + \frac{\partial \bos{\Gamma}(\bos{\gamma}', s; t)}{\partial \tau}
\end{equation}
where we have used $\dot{\tau} = 1$. The term $\frac{\partial \bos{\Gamma}(\bos{\gamma}', s; t)}{\partial \tau}$ can be interpreted as the change in the position in phase-space at time $t$ if the system was initiliazed at time $\tau+d\tau$ instead of $\tau$. Following the same line of arguments as before, we write the equation of motion for the observable $A$, i.e.
\begin{equation}
\label{EOM_A_augm}
\frac{d A(\bos{\gamma}', s; t)}{d t} =  \dot{\bos{\gamma}'}(\bos{\gamma}') \cdot \frac{\partial A(\bos{\gamma}', s; t)}{\partial \bos{\gamma}'}  
\end{equation}
where $A(\bos{\gamma}', t)$ is the value of $A$ at time $t$ for a trajectory initialized at position $\bos{\gamma}'$ in the augmented phase-space. 
We now introduce the modified Liouvillian operator 
\begin{equation}
i\mathcal{L}' = \dot{\bos{\gamma}'}(\bos{\gamma}') \cdot \frac{\partial }{\partial \bos{\gamma}'}  
\end{equation}
This allows to formally integrate the equation of motion for $A(\bos{\gamma}', s; t)$ as 
\begin{equation}
\label{A_augm}
A(\bos{\gamma}', s; t) = e^{i\mathcal{L}'(t-s)}\mathbb{A}(\bos{\gamma}')
\end{equation}
In most concrete situations, the trajectories are all initialized at equivalent reference times, meaning that we start all the configurations with the same value of $\tau$ that one can arbitrarily set to $\tau=s$. If this is the case, with the previous notation we should have 
\begin{equation}
\label{EOM_A_td}
\frac{d A(\bos{\gamma}, s; t)}{d t} =  \dot{\bos{\gamma}}(\bos{\gamma}) \cdot \frac{\partial A(\bos{\gamma}, s; t)}{\partial \bos{\gamma}}   + \lim_{\tau \rightarrow s}  \frac{\partial A\left(\bos{\gamma}', s; t\right)}{\partial \tau}  
\end{equation}
where $A(\bos{\gamma}, s; t) = \lim_{\tau \rightarrow s} A(\bos{\gamma}', s; t)$ has the same meaning as in the previous paragraph, and $\dot{\bos{\gamma}}(\bos{\gamma}) = \lim_{\tau\rightarrow s}\dot{\bos{\gamma}}(\bos{\gamma}')$ is the initial/reference value of the flow field.

Regarding the Schrödinger picture, one might want to write Liouville's theorem, i.e.~the conservation of probability, in the augmented phase-space. Even though eq.~(\ref{liouville_theorem_timedep}) remains valid, it is also possible to write it in a fixed flow field, reading 
\begin{equation}
\label{augm_liouville_theorem}
\frac{\partial \rho(\bos{\gamma}',t)}{\partial t} +  \frac{\partial}{\partial \bos{\gamma}'} \cdot \left[\rho(\bos{\gamma}',t)\dot{\bos{\gamma}'}(\bos{\gamma}',t)\right]= 0
\end{equation}
where $\rho(\bos{\gamma}',t)$ is the probability density in the augmented phase-space at time $t$. Defining a propagator $i\revL'$, in analogy, as
\begin{equation}
\label{def_revL}
i\revL' = \frac{\partial}{\partial \bos{\gamma}'} \cdot \left[ \dot{\bos{\gamma}'}(\bos{\gamma}',t)  \cdots \right]
\end{equation}
Liouville's theorem becomes
\begin{equation}
\partial_{t}\rho(\bos{\gamma'},t) + i\revL'\rho(\bos{\gamma'},t) = 0
\end{equation}
and thus, the propagation of density is integrated as 
\begin{equation}
\label{prop_density_augm}
\rho(\bos{\gamma}',t) = e^{-i\revL'(t-s)}\rho(\bos{\gamma}, s)
\end{equation}

Practically, most experiments are carried out such that all trajectories are initialized in a hyperplane $\tau=s$, in which case the connection with the previous notation is 
\begin{equation}
\label{relation_prime}
\rho (\bos{\gamma}', t) = \rho (\bos{\gamma}, t) \delta(t-\tau)
\end{equation}
The time-dependent picture is now complete, on the Heisenberg level as well as on the Schrödinger level.  (A concrete simple example is shown in the appendix.) In addition, we can establish a relation between these two approaches. Making use of eqn.~(\ref{relation_prime}), (\ref{rho_texp}) and (\ref{prop_density_augm}), we can write
\begin{align}
e^{i\revL'(t-s)} [\delta(s&-\tau)  \rho(\bos{\gamma}, s) ] = \nonumber \\
& \delta(t-\tau) \exp_{+}\left\{ - \int_{s}^{t}i\mathcal{L}_{t'} dt' \right\} \rho(\bos{\gamma}, s)
\end{align}
This equation should be true for any reference phase-space probability density $\rho(\bos{\gamma}, s)$ , thus we have
\begin{equation}
e^{-i\revL'(t-s)}\left[\delta(s-\tau) \boldsymbol{\cdot} \right] = \delta(t-\tau) \exp_{+}\left\{ - \int_{s}^{t}i\mathcal{L}_{t'} dt' \right\}
\end{equation}
which directly connects our approach with the one by Holian and Evans. We can now safely define the time-dependent averages.

\subsection{Time-dependent averages}

To define the averages correctly in the augmented phase-space, we can work directly with  $A(\bos{\gamma}', t)$ itself, and therefore use eqn.~(\ref{A_augm}) together with  eqn.~(\ref{relation_prime}). In the Schrödinger picture, we would have
\begin{align}
\left\langle A(t) \right\rangle &= \int d\mathbf{\bos{\gamma}'} \rho (\bos{\gamma}', t)  \mathbb{A}(\bos{\gamma}') \nonumber \\
 &= \int d\mathbf{\bos{\gamma}'} e^{-i\revL'(t-s)}\left[\rho (\bos{\gamma}, s) \delta(s-\tau)\right]  \mathbb{A}(\bos{\gamma}')
\end{align} 
The integration by parts that is usually performed to switch to the Heisenberg picture is still valid in the augmented phase-space. In fact, $\left( i\revL' \right)^{\dagger} = -i\mathcal{L}'$, therefore we have
\begin{align}
\left\langle A(t) \right\rangle &= \int d\mathbf{\bos{\gamma}'} \rho (\bos{\gamma}, s) \delta(s-\tau)  e^{i\mathcal{L}'(t-s)}\mathbb{A}(\bos{\gamma}') \nonumber \\
&= \int d\mathbf{\bos{\gamma}} \rho (\bos{\gamma}, s) \left[ \int d\tau \delta(s-\tau)  A(\bos{\gamma}', s; t) \right]
\end{align} 
Since we have $A(\bos{\gamma}, s; t) = \lim_{\tau\rightarrow s} A(\bos{\gamma}', s; t) = \int d\tau \delta(s-\tau)  A(\bos{\gamma}', s; t)$, we finally can write
\begin{align}
\left\langle A(t) \right\rangle &= \int d\mathbf{\bos{\gamma}} \rho (\bos{\gamma}, s)  A(\bos{\gamma}, s; t) 
\end{align} 
We recover safely the definition of the time-dependent average of $A$ that one would naturally define in the Heisenberg picture using the usual phase-space.

\section{Conclusion}
In this paper, we have shown how to derive the equation of motion of a coarse-grained observable in the case of a system that is governed by microscopic dynamics that depends explicitly on time, and in the case of an observable having itself an explicit dependence on time. We have shown that the basic steps of the projection operator formalism can still be applied although the propagator of the process is more complicated than a simple exponential operator. We have also derived an improved version of the fluctuation-dissipation-like relation, linking the memory function with the autocorrelation of the fluctuating force between two arbitrary times.

In addition, we have introduced the notion of an augmented phase-space together with a modified Liouvillian operator, which allows us to write an effectively time-independent equation. Although this concept is not strictly necessary to derive the Generalized Langevin Equation in this context, we think it is a useful conceptual tool that allows to understand the problem in geometrical terms. The fact that the streamlines in this space are fixed by construction helps to generalize the Heisenberg equation together with projection operator formalism that are known for conservative systems. The structure of the equation of motion in this space was shown to be identical as the one of the time-independent dynamics, but calculations of averages must be then performed with care. In particular, we have discussed the equivalence between this method and the one introduced by Holian and Evans using time-ordered exponentials.

\section{Acknowldegements}

We thank F.~Schmid, G.~Jung and G.~Amati for useful discussions.  This project has
been financially supported by the National Research Fund Luxembourg (FNR) within the AFR-PhD programme. 

\clearpage

\section{Supplemental material}

\subsection{Detailed derivation of the Heisenberg equation in time-independent dynamics}

We show in this appendix a geometric derivation of the Heisenberg equation. We use the same notations as in the main text.

\begin{figure}
\begin{center}
\includegraphics[width=\linewidth]{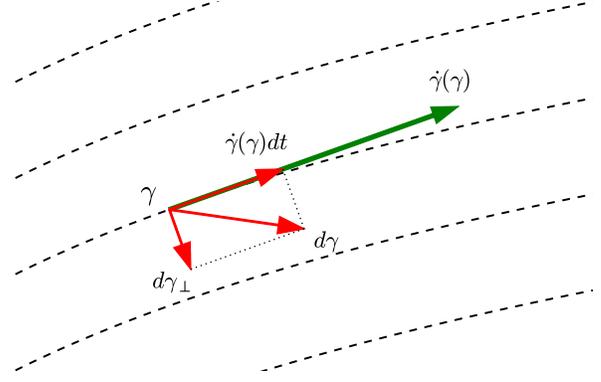}
\end{center}
\caption{Sketch of the decomposition (\ref{decomposition}). The dotted lines represent streamlines in phase-space.}
\label{dgamma}
\end{figure}

Let us first consider an infinitesimal displacement $d\bos{\gamma}$ in phase-space, proportional to $\dot{\bos{\gamma}}(\bos{\gamma}, t)$, i.e. $d\bos{\gamma} = \dot{\bos{\gamma}}(\bos{\gamma}, t)dt$. In the case of a time-independent flow field $\dot{\bos{\gamma}}(\bos{\gamma}, t) = \dot{\bos{\gamma}}(\bos{\gamma})$, that is for a Hamiltonian that does not depend on time, we have  
\begin{equation}
\label{disp_along_supp}
\bos{\Gamma}(\bos{\gamma}+\dot{\bos{\gamma}}(\bos{\gamma})dt, s; t) = \bos{\Gamma}(\bos{\gamma}, s; t+dt)
\end{equation}
Now, let us consider an arbitrary displacement $d\bos{\gamma}$ that we decompose into its component along $\dot{\bos{\gamma}}(\bos{\gamma})$ and its other components in the orthogonal subspace, i.e. 
\begin{equation}
\label{decomposition_supp}
d\bos{\gamma} = \dot{\bos{\gamma}}(\bos{\gamma})dt + d\bos{\gamma}_{\perp}
\end{equation}
where $d\bos{\gamma}_{\perp}\cdot \dot{\bos{\gamma}} = 0$, and therefore $dt = d\bos{\gamma} \cdot \dot{\bos{\gamma}} / |\dot{\bos{\gamma}}|^{2}$. (A sketch of this is shown in figure \ref{dgamma}.) We use this decomposition to write
\begin{align}
&\bos{\Gamma}(\bos{\gamma}+d\bos{\gamma}, s; t) = \bos{\Gamma}\left( \bos{\gamma}+ d\bos{\gamma}_{\perp} +\dot{\bos{\gamma}}(\bos{\gamma})dt , s; t \right) \nonumber \\
&= \bos{\Gamma}\left(\bos{\gamma}+ d\bos{\gamma}_{\perp} +\dot{\bos{\gamma}}(\bos{\gamma}+d\bos{\gamma}_{\perp})dt - \frac{\partial \dot{\bos{\gamma}}}{\partial \bos{\gamma}}(\bos{\gamma})\cdot d\bos{\gamma}_{\perp}dt, s; t \right)
\end{align}
where we have used $\dot{\bos{\gamma}}(\bos{\gamma} + d\bos{\gamma}_{\perp}) = \dot{\bos{\gamma}}(\bos{\gamma}) +\frac{\partial \dot{\bos{\gamma}}}{\partial \bos{\gamma}}(\bos{\gamma})\cdot d\bos{\gamma}_{\perp}$. We neglect the second-order term in $d\bos{\gamma}_{\perp}dt$ and we use eqn.~(\ref{disp_along}) to find
\begin{equation}
\bos{\Gamma}(\bos{\gamma}+d\bos{\gamma}, s; t)
= \bos{\Gamma}(\bos{\gamma}+ d\bos{\gamma}_{\perp}, s; t+dt)
\end{equation}
We finally expand this as
\begin{equation}
\bos{\Gamma}(\bos{\gamma}+d\bos{\gamma}, s; t)
= \bos{\Gamma}(\bos{\gamma} , s; t) + \frac{\partial \bos{\Gamma}(\bos{\gamma}, s; t)}{\partial \bos{\gamma}} \cdot  d\bos{\gamma}_{\perp} + \frac{\partial \bos{\Gamma}(\bos{\gamma}, s; t)}{\partial t}  dt
\end{equation}
where we have again neglected the second-order term in $d\bos{\gamma}_{\perp}dt$. We now replace $d\bos{\gamma}_{\perp}$ by $d\bos{\gamma} - \dot{\bos{\gamma}}(\bos{\gamma})dt$ and we use $\bos{\Gamma}(\bos{\gamma}+d\bos{\gamma}, s; t) = \bos{\Gamma}(\bos{\gamma} , s; t) + \frac{\partial \bos{\Gamma}}{\partial \bos{\gamma}}(\bos{\gamma}, s; t) \cdot  d\bos{\gamma}$ to finally obtain
\begin{equation}
\label{jacobian_supp}
\frac{d \bos{\Gamma}(\bos{\gamma}, s; t)}{d t} =  \frac{\partial \bos{\Gamma}(\bos{\gamma}, s; t)}{\partial \bos{\gamma}} \cdot  \dot{\bos{\gamma}}(\bos{\gamma}) 
\end{equation}
We can now use eqn.~(\ref{Gamma_gammadot}) and insert the result above into eqn.~(\ref{partial_A_heisenberg}). Explicitly noting all the indices we obtain
\begin{equation}
\frac{d A(\bos{\gamma}, s; t)}{d t} =  \sum_{i,j}\dot{\gamma}_{i}(\bos{\gamma}) \frac{\partial \Gamma_{j}(\bos{\gamma}, s; t)}{\partial \gamma_{i}}   \frac{\partial \mathbb{A}\left(\bos{\Gamma}(\bos{\gamma}, s; t) \right)}{\partial \gamma_{j}}
\end{equation}
This can be rewritten as 
\begin{equation}
\frac{d A(\bos{\gamma}, s; t)}{d t} =  \sum_{i}\dot{\gamma}_{i}(\bos{\gamma})  \frac{\partial }{\partial \gamma_{i}} \left[ \mathbb{A} \left(\bos{\Gamma}(\bos{\gamma}, s; t) \right) \right]
\end{equation}
By making use of the correspondance $A(\bos{\gamma},s; t) = \mathbb{A}(\bos{\Gamma}\left( \bos{\gamma},s; t \right))$, one can finally rewrite
\begin{equation}
\frac{dA(\bos{\gamma}, s; t)}{dt} =  \dot{\bos{\gamma}}(\bos{\gamma}) \cdot \frac{\partial A(\bos{\gamma}, s; t)}{\partial \bos{\gamma}}  = i\mathcal{L}A(\bos{\gamma},s; t)
\end{equation}
This equation is the expected Heisenberg equation. However, the derivation made use of eqn.~(\ref{disp_along}), which is valid only in the case of a flow field that does not depend on time. Therefore, this approach cannot work if the Liouvillian has an explicit dependence on time.

\subsection{Simple example : the harmonic oscillator}
To illustrate the discussion of the concept of augmented phase-space, we study a simple 1-particle harmonic oscillator, with the Hamiltonian
\begin{equation}
\mathcal{H}_{t}(p,x) = \frac{p^{2}}{2m} + \frac{k(t)x^{2}}{2}
\end{equation}
We want to show in this part how eqn.~(\ref{augm_jacobian}) is verified, first in a case of a constant spring constant, and then with a time-dependence. 

In the case of a time-independent Hamiltonian, i.e. $k(t) = k_{0}$, the connection with the notation defined in the main text of this paper is as follows 
\begin{align}
\bos{\gamma} &= (x,p) \\
\dot{\bos{\gamma}}(\bos{\gamma}, t) &= (\dot{x}, \dot{p}) = (p/m, -k_{0}x) \\
\bos{\Gamma}\left( \bos{\gamma}, s; t \right) &= (X(\bos{\gamma}, s; t), P(\bos{\gamma}, s; t)) 
\end{align}
where $X(\bos{\gamma}, s; t)$ and $P(\bos{\gamma}, s; t)$ are respectively the position and momentum of the particle at time $t$ is the system was located at the position $x$ and momentum $p$ at time $s$. We will now forget about the time $s$ that we set to $s=0$, i.e. $\bos{\Gamma}\left( \bos{\gamma}, s; t \right)$ becomes $\bos{\Gamma}\left( \bos{\gamma},  t \right)$. 

The motion of the particle in phase-space can be analytically solved, reading
\begin{align}
X(\bos{\gamma},t) &= x\cos(\omega t) + \frac{p}{m \omega }\sin(\omega t) \\
P(\bos{\gamma},t) &= -m\omega x\sin(\omega t) + p\cos(\omega t)
\end{align}
with $\omega = \sqrt{k_{0}/m}$. Using this notation, we compute the Jacobian matrix $\frac{\partial \bos{\Gamma}}{\partial \bos{\gamma}}(\bos{\gamma}, t)$, and obtain
\begin{equation}
\frac{\partial \bos{\Gamma}}{\partial \bos{\gamma}}(\bos{\gamma}, t) \cdot \dot{\bos{\gamma}}(\bos{\gamma}) = \left( 
\begin{tabular}{c }
$\frac{p\cos(\omega t)}{m} - \frac{k_{0} x\sin(\omega t)}{m\omega}$  \\
$-k_{0}x\cos(\omega t) - \omega p \sin(\omega t)$
\end{tabular}
\right)
\end{equation}
On the other hand, we have
\begin{equation}
\dot{\bos{\gamma}}(\bos{\Gamma}(\bos{\gamma}, t)) = \left( 
\begin{tabular}{c }
$\frac{p\cos(\omega t)}{m} - \omega x\sin(\omega t)$  \\
$ - \frac{kp \sin(\omega t)}{m\omega} - kx\cos(\omega t)$
\end{tabular}
\right)
\end{equation}
Using again $\omega = \sqrt{k_{0}/m}$, the identity (\ref{jacobian_supp}) is verified, and therefore eqn.~(\ref{augm_jacobian}) is also verified since both equations are identical in the case of time-independent Hamiltonian. 

Now, we choose a time-dependent trap, i.e.~$k(t) = k_{0}e^{-\Omega t}$, and we want to verify again eqn.~(\ref{augm_jacobian}). As explained earlier, the dependence on $\tau$ can be introduced by solving the equation of motion for a shifted Hamiltonian $\tilde{\mathcal{H}}_{t} = \mathcal{H}_{t+\tau}$. Thus, we solve the equation 
\begin{equation}
m\ddot{Z}(t) + k(t+\tau)Z(t) = 0
\end{equation}
with the initial conditions $Z(0) = x$ and $Z'(0) = p/m$. The solution can be expressed analitycally using Bessel functions, i.e.
\begin{widetext}
	\begin{equation}
	X(\bos{\gamma}',t) = \frac{ J_{0}(\alpha(t+\tau)) \left[ \omega X Y_{1}(\alpha(\tau)) - PY_{0}(\alpha(\tau))e^{\Omega \tau/2} \right] - Y_{0}(\alpha(t+\tau))  \left[ \omega X J_{1}(\alpha(\tau)) - PJ_{0}(\alpha(\tau))e^{\Omega \tau/2} \right]}{\omega \left[J_{0}(\alpha(\tau)) Y_{1}(\alpha(\tau)) - Y_{0}(\alpha(\tau)) J_{1}(\alpha(\tau))\right]}
	\end{equation}
\end{widetext}
with $\alpha(t) = 2\frac{\omega}{\Omega}e^{-\Omega t/2 }$, $J_{n}$ is the Bessel function of first kind and $Y_{n}$ is the Bessel function of second kind. With this result, one can compute $P(\bos{\gamma}',t)$ and also compute the derivatives of these two functions with respect to either $x$, $p$ or $\tau$, and we also have $\dot{\bos{\gamma}}'(\bos{\gamma}') = (p/m, -k(\tau)x, 1)$. We have everything to verify eq.(\ref{augm_jacobian}), which turns out to be satisfied.
\begin{center}
	\begin{figure}
		\includegraphics[width=\linewidth]{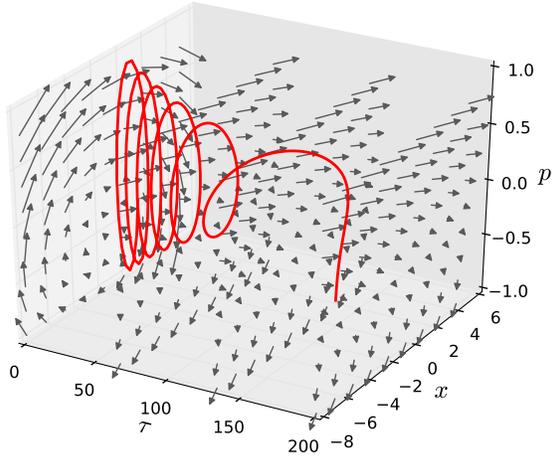}
		\caption{We plot here the two trajectory in the augmented phase-space, for the particular choice $x=1$, $p=0$ and $\tau=0$ for the initial state in the augmented phase-space, and $k_{0}=m=1$, $\Omega =0.05$. The gray arrows represent the vectors $\bos{\gamma}'(\bos{\gamma})$.}
		\label{verif}
	\end{figure}
\end{center}

\end{document}